# Electric-field-induced strain-mediated magnetoelectric effect in CoFeB-MgO magnetic tunnel junctions

V. B. Naik,[1,*] H. Meng,[1] J. X. Xiao,[2] R. S. Liu,[1,†] A. Kumar,[2] K. Y. Zeng,[2] P. Luo[1] and S. Yap[1]

[1]Data Storage Institute, A*STAR (Agency for Science Technology and Research), 5 Engineering Drive 1, DSI Building, Singapore 117608, Singapore.

[2]Department of Mechanical Engineering, National University of Singapore, 9 Engineering Drive 1, Singapore 117576, Singapore.

**Abstract**

**Magnetoelectric coupling between magnetic and electric dipoles is one of the cornerstones of modern physics towards developing the most energy-efficient magnetic data storage. Conventionally, magnetoelectric coupling is achieved in single-phase multiferroics or in magnetoelectric composite nanostructures consisting of ferromagnetic and ferroelectric/piezoelectric materials. Here, we demonstrate an electric-field-induced strain-mediated magnetoelectric effect in ultrathin CoFeB/MgO magnetic tunnel junction employing non-piezoelectric material, which is a vitally important structure for spintronic devices, by using dynamical magnetoelectric and piezoresponse force microscopy measurement techniques. We show that the applied electric-field induces strain in a few atomic layers of dielectric MgO which is transferred to magnetostrictive CoFeB layer, resulting in a magnetoelectric effect of magnitude up to 80.8 V cm$^{-1}$ Oe$^{-1}$ under -0.5 V. The demonstrated strain-mediated magnetoelectric effect with an electric field in magnetic tunnel junctions is a significant step towards exploring magnetoelectrically controlled spintronic devices for low-power and high density magnetic data storage applications.**

---

[*] Email: Vinayak_BN@dsi.a-star.edu.sg
[†] Current address: Western Digital Corp., 44100 Osgood Road, Fremont, CA 94539, US.

The magnetoelectric (ME) effect in a solid—that is, the induction of a magnetization by means of an electric field (*E*-field) and the induction of an electric polarization by means of a magnetic field, is an exciting new area of condensed matter research towards developing low-power spintronics technology, such as magnetic random access memory (MRAM), magnetic sensors and spin logic devices[1,2,3,4,5,6,7,8]. Exploiting ME effect in spintronics technology is a promising way to control magnetization with *E*-field rather than with electric current or magnetic field in order to avoid the problems of heat dissipation and stray magnetic field. Conventionally, the ME effect is observed in single-phase multiferroics, which can exhibit simultaneously electric and magnetic orders[9,10,11,12,13] or in ME composite nanostructures via strain-mediated cross interaction between the magnetostrictive effect in the ferromagnetic phase and piezoelectric effect in the ferroelectric/piezoelectric phase[14,6,15,16,17,18,19]. Since the single phase multiferroic materials with a high inherent coupling between the multiferroic order parameters especially above room temperature are rare and the ME effects are typically too small, the ME composite nanostructures have drawn significant interest in recent years due to their much higher ME coefficient and their much higher operation temperature. However, the integration of such ME multi-phase nanostructures with conventional magnetic tunnel junctions (MTJs) using spintronic process technology remains one of the challenging issues due to incompatibility of material processes. In addition, maintaining the stability of ferroelectricity/piezoelectricity at the very small thickness in order to achieve low resistance-area product for complementary metal-oxide–semiconductor circuit compatibility is also extremely difficult. Therefore, the exploration of ME effect in conventional and most established material system without compromising other spintronics properties is of primary importance.

For spintronics technology, MTJs based on usual dielectric materials such as Al-oxide, MgO etc. have been widely studied[20,21,22,23,24], and particularly MgO MTJs have attracted the interest of researchers worldwide in recent years because of its very high tunnel magnetoresistance (TMR) ratio and high thermal stability[25,26]. However, the $E$-field-induced ME effect in such MTJs employing non-piezoelectric materials has not been reported yet. In this work, we demonstrate the ME effect in voltage-biased ultrathin CoFeB/MgO MTJ by employing dynamical ME and piezoresponse force microscopy (PFM) measurements. We find that the applied $E$-field induces strain in a few atomic layers of dielectric MgO, which is transferred to the magnetostrictive CoFeB layer to achieve a strong ME coupling between MgO and CoFeB layers. Our results of such $E$-field-induced strain-mediated ME coupling will shed light on understanding the recently observed $E$-field-controlled perpendicular magnetic anisotropy (PMA) properties in CoFeB/MgO MTJs[27,28,29,30] and may accelerate the development of energy efficient and high density magnetic data storage.

MTJ stacks were deposited on thermally oxidized Si substrate at room temperature. The investigated structure is made of Si/SiO$_2$/bottom electrode/Ta 2/CoFeB 1/MgO 1.8/CoFeB $t$ (thickness) = 1.2-1.7/Ta 5/Ru 10 (numbers are nominal thicknesses in nanometres), where the composition of CoFeB is Co$_{40}$Fe$_{40}$B$_{20}$ (atomic percentage). The multilayer was processed into circular MTJ devices with a diameter of 85 nm by electron beam lithography (EBL) and ion beam etching processes, and annealed in a vacuum oven at 300 °C for 1 hour. The magnetoresistance loops were measured at room temperature with a direct current (d.c.) voltage ($V_{DC}$) under a perpendicular scanning magnetic field ($H_{DC}$). Here, the positive bias voltage is defined as inducing the tunnelling of electrons from the bottom CoFeB electrode to the top CoFeB electrode.

Figure 1a is a schematic drawing of a MTJ device under external *E*-field. Fig. 1b shows the configuration of *E*-field-induced ME voltage measurement setup, where we used a dynamic lock-in technique to measure the alternating current (a.c.) ME voltage ($V_{ME}$) across MTJ device due to presence of a small a.c. magnetic field ($H_{AC}$) with frequency (*f*) under applied d.c. voltage ($V_{DC}$) as a function of perpendicular $H_{DC}$ (see Methods section). Fig. 1c shows the junction resistance (*R*) of a device with top CoFeB free layer (FL) of thickness 1.2 nm as a function of $H_{DC}$ at $V_{DC}$ = -0.5 V. The FL magnetization switches smoothly for *t* = 1.2 nm, but it switches sharply for higher values *t*, identified through a thickness dependent study (see Supplementary Information). Fig. 1d shows $V_{ME}$ as a function of $H_{DC}$ measured at the optimum frequency of 933 Hz under $H_{AC}$ = 1.3 Oe for two voltage bias values, $V_{DC}$ = 0 and -0.5 V. A sharp peak is resolved in $V_{ME}$ around the magnetization switching field of FL at $V_{DC}$ = -0.5 V while no peak appears at $V_{DC}$ = 0 V. Generally, this kind of peak in ME voltage is observed in conventional ME composite materials that originates from the magnetic-mechanical-electric coupling through the strain transfer across the interface, which is attributed to a sharp change in piezomagnetic coefficient, $q = d\lambda/dH$, where $\lambda$ is the magnetostriction of magnetostrictive layer in the composite system[19]. Therefore, given that the CoFeB is magnetostrictive with $\lambda_s$ = 2 x 10$^{-5}$ (where $\lambda_s$ is the saturation magnetostriction)[31], a sharp peak in $V_{ME}$ in our CoFeB/MgO MTJ device under *E*-field indicates the *E*-field-induced ME coupling between the magnetostrictive CoFeB and dielectric MgO layers through strain transfer under the influence of *E*-field at nanoscale.

The evolution of $V_{ME}$ as a function of negative bias voltages ($V_{DC}$ = -0.5 V to 0 V) and $H_{DC}$ is shown in Fig. 2a, where the $H_{DC}$ is swept from 0 to -300 Oe with the pinned bottom CoFeB reference layer (RL) magnetization pointing upwards (See Supplementary Information for *E*-field dependent minor TMR loops). Clearly there is no sign of any peak in $V_{ME}$ when $V_{DC} = 0$ (no $E$-field). However, as the magnitude of $V_{DC}$ ($E$-field) increases, a peak emerges in $V_{ME}$, and its magnitude also increases with increasing $V_{DC}$. The inset shows the map of $V_{ME}$ as a function of $V_{DC}$ and $H_{DC}$, where the magnitudes of the peaks under positive bias voltages are slightly lower than that of negative bias voltages. The maximum ME coefficient value of 36.4 V cm$^{-1}$ Oe$^{-1}$ is obtained for $t = 1.2$ nm under $V_{DC} = -0.5$ V.

A plot of magnitude of $V_{ME}$ as a function of $f$ measured at fixed $H_{AC} = 1.3$ Oe and $V_{DC} = -0.7$ V is shown in Fig. 2b. The value of $V_{ME}$ increases rapidly and shows a peak around the resonance frequency ($f_r$) of 933 Hz and then decreases monotonically for $f > f_r$. The $f_r$ value found in this CoFeB/MgO system is close to the reported value of $f_r = 1197$ Hz in CoFeBSi-based ME compound composed of a known piezoelectric material, AlN[32]. The sensitivity of this $E$-field-induced ME effect is shown in Fig. 2c, where $V_{ME}$ measured as a function of $H_{AC}$ at $f_r = 933$ Hz and $V_{DC} = -0.7$ V. This $E$-field-induced ME effect is detectable even in the small a.c. magnetic field of 0.06 Oe, and its magnitude is linearly increasing with increasing value of $H_{AC}$, resembling the behavior of a ME sensor made of perfect magnetostrictive and piezoelectric layers.

Although Gopal and Spaldin theoretically predicted[33] that the structurally relaxed MgO can show piezoelectricity with piezoelectric constant as high as $e_{33} = 2.26$ C/m$^2$, there has been no experimental evidence of piezoelectric effect in nanoscaled MgO material so far. In ultrathin films of (Fe, Co, CoFeB)/MgO, the ferromagnet/dielectric interface plays a major role as compare to the bulk effect. At the interface, the atomic orbits of metal and O atoms hybridize and form metal-O bonds that results in elastic distortion of the crystal lattice of the ferromagnet and MgO, which can induce a tensile strain[34,35,36]. It has been proposed that the applied $E$-field can displace O atom at the metal–O interface in high ionic mobility oxides[37]. Recently, using molecular dynamics simulation, Enyashin *et. al.*,[38] showed that the structure of nano-scaled MgO can be changed from cubic phase to non-cubic phase by applying a mechanical strain. To corroborate our understanding that the observed experimental results in our CoFeB/MgO structure is indeed originated from the *E*-field-induced strain-meditated ME coupling, we have investigated the piezoelectric effect in nanopillar device with 5 nm MgO using a commercial vertical PFM (See Methods section). Fig. 3a and 3b show PFM images of the device at applied a.c. voltage amplitude of 1 V and 2 V, respectively. Both images show a clear color contrast between the circled and the rest areas with an enhancement for the higher applied voltage, indicating the development of strain under applied voltage in nanoscaled MgO and it becomes stronger as the voltage increases. In Fig. 3c, we show the strain coefficient values ($d_{zz}$) at several voltage amplitudes and an approximately linear dependence of $d_{zz}$ on voltage can be observed. Meanwhile, the dependence of peak value of $V_{ME}$ as a function of $V_{DC}$ is shown in Fig. 3d for comparison, which depicts the linear dependence of $V_{ME}$ for both polarities of $V_{DC}$. The increasing magnitude of both $V_{ME}$ and $d_{zz}$ with increasing *E*-field confirms that the applied *E*-field induces a strong ME coupling in CoFeB/MgO MTJ through strain transfer across the CoFeB/MgO interface via the cross interaction between the magnetostrictive effect (mechanical/magnetic) in CoFeB and the piezoelectric effect (electrical/mechanical) in MgO at nanoscale.

Interestingly, we found that the ME voltage is highly dependent on the angle $\phi$ between the FL magnetization direction and the applied $H_{DC}$ due to applied *E*-field. Here, we illustrate the phenomenon by choosing thick (*t* = 1.6 and 1.7, the numbers are nominal thicknesses in nanometres) CoFeB FL such that its magnetization has a crossover point of PMA to in-plane (see Supplementary Information). Fig. 4a and 4c show the normalized TMR loops measured at 3 selected $V_{DC}$ values for *t* = 1.6 nm and 1.7 nm, respectively. A sharp switching in TMR loop at $V_{DC}$ = 50 mV for *t* = 1.6 nm indicates that the FL magnetization has strong PMA (Fig. 4a). The PMA of FL becomes stronger under positive bias voltage ($V_{DC}$ = +1 V) while it reduces under negative bias voltage ($V_{DC}$ = -1 V). Fig. 4b shows the $V_{ME}$ evolution as a function of $V_{DC}$ and $H_{DC}$ at the optimal frequency of 2733 Hz under $H_{AC}$ = 1.3 Oe. The ME voltage signal is observed under negative $V_{DC}$ (-$V_{DC}$ > -0.2 V) while no ME voltage under positive $V_{DC}$. The absence of ME voltage under positive bias voltage is a sign of no magnetostriction as the magnetization easy axis is parallel to the direction of applied $H_{DC}$, which is in agreement with the fact that the ME voltage depends on the value of $\lambda$ such that $\lambda$ = 0 for $\phi$ = 0 and $\lambda = \lambda_s$ for optimum value of $\phi$[39]. A maximum ME coefficient of 80.8 V cm$^{-1}$ Oe$^{-1}$ is obtained for $t$ = 1.6 nm at $f_r$ under $V_{DC}$ = -0.5 V.

The normalized TMR minor loops for $t$ = 1.7 nm at $V_{DC}$ = 50 mV is not a sharp rectangular shape (Fig. 4c), indicating that the magnetization of FL is tilted which can be clearly seen in the inset where a full TMR loop is shown. The shape of the minor loop changes significantly as the applied $V_{DC}$ changed from -1 V to 1 V, which is due to the tuning of $\phi$ by $V_{DC}$ i.e., the negative $V_{DC}$ increases the value of $\phi$, while the positive $V_{DC}$ decreases the value of $\phi$. Note that the magnetization of FL has not attained a complete PMA i.e., its magnetization is slightly tilted, even under the maximum positive bias voltage applied ($V_{DC}$ = 1 V). In contrast to that in Fig 4b, the evolution of $V_{ME}$ is observed only under positive $V_{DC}$ values for $t$ = 1.7 nm and there is no peak in $V_{ME}$ under negative $V_{DC}$, indicating that the value of $\lambda$ is almost negligible for such values of $\phi$ (Fig. 4d).

Finally, we have investigated the effect of d.c. *E*-field polarity on the piezoresponse of nanoscaled MgO. PFM images were collected for a series of d.c. bias voltages ranging from -1.5 V to 1.5 V applied to a nanopillar (see Methods section). Fig. 5a, 5b and 5c show the PFM images at $V_{DC}$ = 0 V, -1.5 V and +1.5 V, respectively. A distinct color contrast in the circular areas can be observed for applied voltage of different polarities, which is due to the effect of *E*-field polarity dependent induced-strain *i.e.*, the negative *E*-field induces a compressive strain (darker color) while the positive *E*-field induces a tensile strain (brighter) in nanoscaled MgO. In Fig. 5d, we plot the $d_{zz}$ as a function of $V_{DC}$ which depicts the linear and polarity dependences of $V_{DC}$, which further confirms that the nanoscaled MgO behaves like a conventional piezoelectric material.

The *E*-field dependent strain in CoFeB/MgO MTJs could have a major role in altering the anisotropy field ($H_k$) of ultrathin CoFeB magnetic layer through $H_k \propto \lambda\sigma$, where $\sigma$ is the stress on the magnetic layer due to strain transfer from nanoscaled MgO. Unlike in the case of magnetic field and spin-transfer-torque switching of FL in MTJ, the magnetic anisotropy energy or barrier height ($E_b$) of a PMA MTJ ($E_b = \frac{1}{2}M_s H_k V$, where $M_s$ − saturation magnetization and $V$ − volume of the magnetic cell) is momentarily lowered under *E*-field. Since the charge density of a metal can be varied only slightly by *E*-field, the change in interfacial magnetic anisotropy energy due to the change in $M_S$ is very small[40]. Therefore, the momentary reduction in $E_b$ may be primarily attributed to the change in $H_k$ under the influence of *E*-field. Indeed, the tuning of $H_k$ by applying *E*-field is a common scheme in ME nanostructures which exhibit a ME coupling between ferromagnetic/piezoelectric layers[14,19]. Recently, it has been shown that the *E*-field can significantly change the $H_k$ of CoFeB layer when it is integrated with a conventional piezoelectric material due to ME coupling between the magnetostrictive CoFeB and piezoelectric layers[41,42]. The change of $H_k$ under *E*-field due to ME coupling relies on the fact that the positive *E*-field strengthens the magnetization along the easy axis and the negative *E*-field tilts the magnetization from the easy axis. Both of these features are associated with the compressive and tensile strain transfer from the piezoelectric layer to a magnetostrictive layer. This kind of *E*-field polarity dependent $H_k$ change is consistent with what we observed in *E*-field dependent PMA change in CoFeB/MgO MTJs as presented in Fig. 4a and 4c, and also in other reports[27,28,30]. Therefore, the *E*-field-induced strain-mediated ME coupling in the confined CoFeB/MgO MTJ stack could contribute to the *E*-field-induced PMA change.

In summary, we have presented a direct evidence for *E*-field-induced ME effect in ultrathin CoFeB/MgO MTJs employing non-piezoelectric material. Using dynamical ME and PFM measurements, we have demonstrated that the observed ME effect is a consequence of ME coupling through *E*-field-induced strain transfer across the CoFeB/MgO interface via the cross interaction between the magnetostrictive effect (mechanical/magnetic) in CoFeB layer and the piezoelectric effect (electrical/mechanical) in MgO layer at nanoscale. We obtained a maximum ME coefficient value of 80.8 V cm$^{-1}$ Oe$^{-1}$ under an applied d.c. voltage of -0.5 V. Such *E*-field-induced strain-mediated ME effect in MTJs represents a new avenue for exploring magnetoelectrically controlled spintronic devices so as to develop more energy-efficient spintronic devices.

## Methods:

A dynamic lock-in method is employed to measure the ME properties. Fig. 1b shows the schematic diagram of the setup used to measure the *E*-field-induced a.c. ME voltage ($V_{ME}$) across the MTJ device, which is similar to the experimental setup described by Fuentes-Cobas et al.[43] An electromagnet was used to generate the perpendicular d.c. magnetic field ($H_{DC}$). The d.c. *E*-field ($V_{DC}$) is applied by using a Keithley's 2602 source meter. Keithley's 6221 current source was used to drive a pair of Helmholtz coils to generate a.c. magnetic field ($H_{AC}$) of amplitude 0-1.3 Oe in the frequency range of $f$ = 0-5 kHz. An odd value of a.c. magnetic field frequency was chosen for measurements in order to avoid the noise signal which arises from the power line frequency (50 Hz) and also its harmonics. In our measurement configuration, both $H_{AC}$ and $H_{DC}$ are parallel to each other and are perpendicular to the plane of MTJ multilayers. The $V_{ME}$ developed across a MTJ device was measured by using a lock-in amplifier (SR830, Stanford Research Systems) as a function of $H_{DC}$. The value of $V_{ME}$ voltage is presented in V cm$^{-1}$ Oe$^{-1}$ using the formula

$V_{ME} = \dfrac{V_o}{\left[(t_{MgO} + t_{CoFeB}) \times H_{AC}\right]}$, where $V_o$ is the voltage measured by lock-in at its input terminal, and $t_{MgO}$ and $t_{CoFeB}$ are the thicknesses of the MgO and CoFeB layers.

For the PFM study, a commercial scanning probe microscopy (SPM) system (MFP-3D, Asylum Research, USA) was used with a commercial software platform (IGOR PRO 6.12A) and SPM control software (Asylum Research, version 090909-1214). A multilayer stack with bottom electrode/Ta 5/CoFeB 1.2/MgO 5/Ta 5/Ru 5 (numbers are nominal thicknesses in nanometres) was deposited on oxidized Si substrate. The multilayer was patterned into circular nanopillars with a diameter of 800 nm by EBL and ion beam etching processes, and annealed in a vacuum oven at 300 °C for 1 hour. To quantitatively characterize the electromechanical properties, the dual AC resonance tracking (DART) PFM mode, which oscillates the conductive cantilever with two frequencies near the contact resonance frequency, was used. An a.c. voltage ranged from 1-2.5 V was applied between the tip and the nanopillar device to determine the strain coefficient ($d_{zz}$) versus $V_{AC}$. In order to study the effect of d.c. bias voltage ($V_{DC}$) on $d_{zz}$, a series of d.c. bias voltages ranged from $V_{DC}$ = -1.5 V to 1.5 V, and an a.c. voltage of $V_{AC}$ = 0.3 V were applied simultaneously to the tip by connecting the device bottom electrode to ground. The value of $d_{zz}$ as a function of $V_{DC}$ is determined by setting the $d_{zz}$ of the unbiased ($V_{DC}$ = 0) region to 0 pm. Here, the positive and negative $d_{zz}$ values indicate the expansion (tensile strain) and contraction (compressive strain) of the material. In all PFM measurements, the Pt-coated Si tips (AC240TM, Olympus, Japan) with average tip radius of 15 nm, nominal stiffness of 2 N/m and resonance frequency of 70 kHz were used. The line scan frequency used was 1 Hz.

**Author contributions**

VBN initiated the study and designed the experiment. VBN, MH, RSL planned the experiment, fabricated the MTJ devices with the help of PL and SY, and performed the experiment and discussed the results. VBN, JXX, AK and KYZ planned and performed PFM experiments and analyzed PFM data. VBN wrote the manuscript with input from MH, JXX and RSL.

**Acknowledgements**

We would like to thank F. Ernult, R. Sbiaa, Y. Yang, J. Shanmugam and S. L. G. Ng for their technical support and useful discussions.

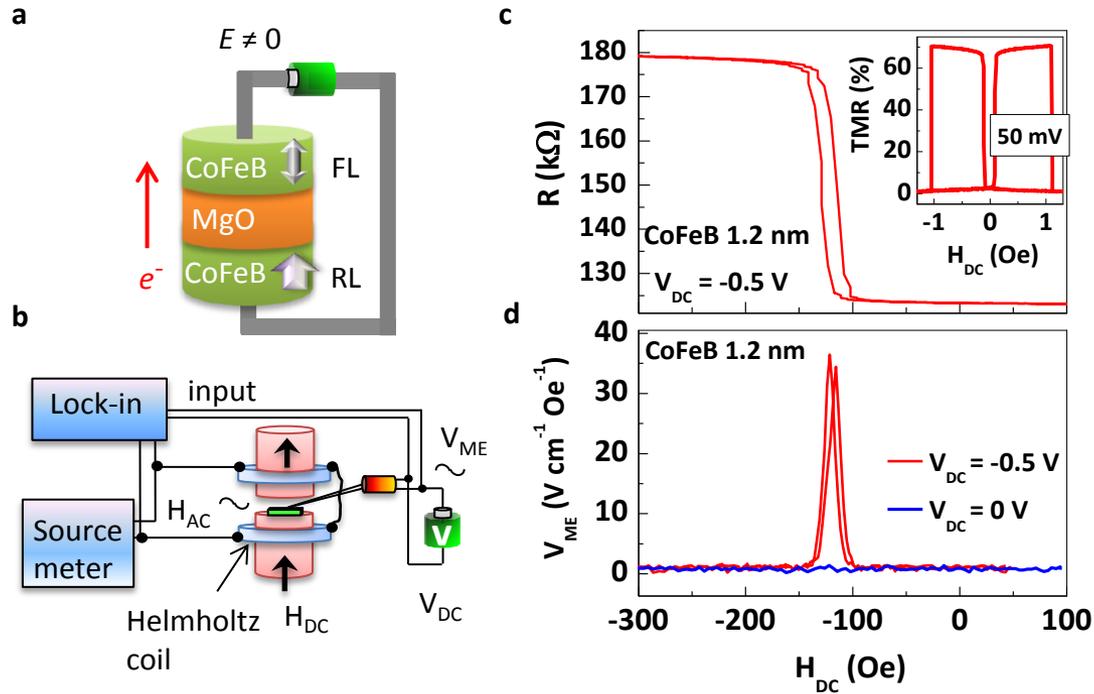

**Figure 1 | Electric-field-induced magnetoelectric effect in CoFeB/MgO/CoFeB MTJ device with interfacial PMA.**
**a**, Schematic drawing of a MTJ device under external *E*-field. The arrow indicates the flow of electrons from bottom CoFeB to top CoFeB.
**b**, Dynamic lock-in measurement setup to measure the *E*-field-induced ME voltage in MTJ device due to presence of a small a.c. magnetic field ($H_{AC}$) and d.c. voltage ($V_{DC}$) under the perpendicular d.c. bias magnetic field ($H_{DC}$). **c**, Minor RH loop of a device at $V_{DC}$ = -0.5 V, and the inset shows the full TMR loop at $V_{DC}$ = 50 mV with a TMR value of 70%. **d**, *E*-field induced ME voltage ($V_{ME}$) as a function of $H_{DC}$ at $V_{DC}$ = 0 and -0.5 V under $H_{AC}$ = 1.3 Oe for *f* = 933 Hz.

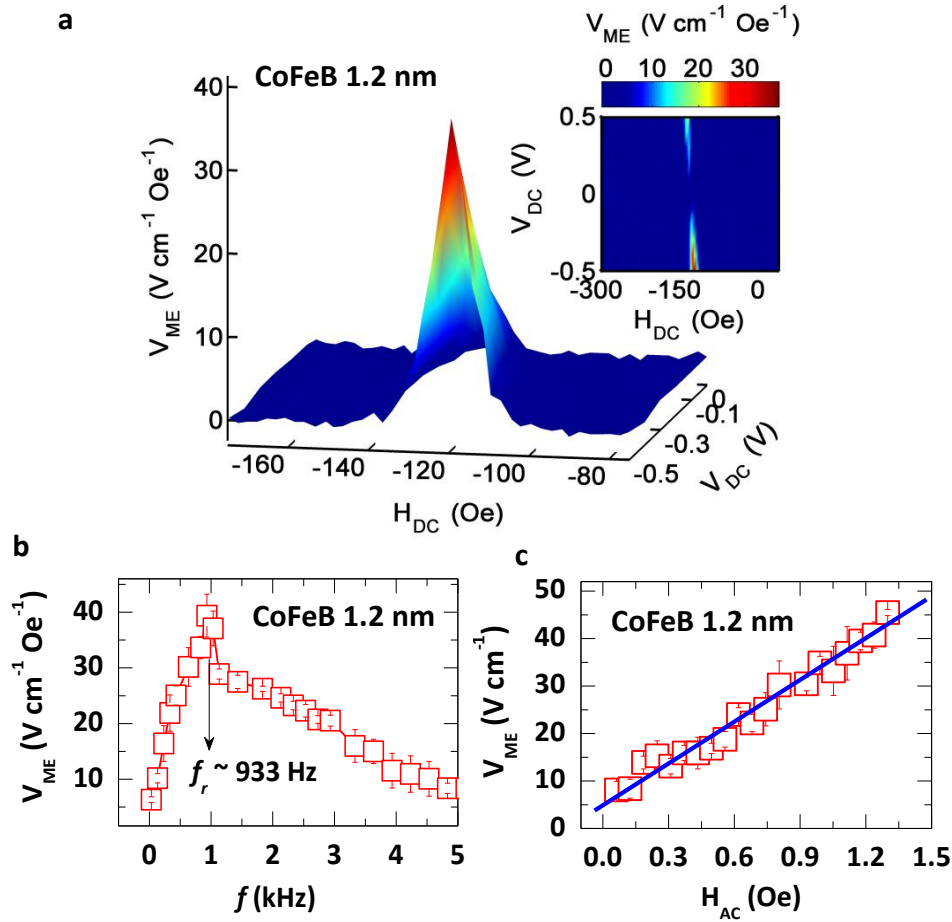

**Figure 2 | Evolution of ME voltage in CoFeB/MgO/CoFeB MTJ device under applied *E*-field.**
**a**, The applied d.c. bias and d.c. magnetic field map of ME voltage for different $V_{DC}$ values from 0 to -0.5 V. The inset shows the map of $V_{DC}$ and $H_{DC}$ of $V_{ME}$ for $V_{DC}$ = -0.5 V to +0.5 V. **b**, The maximum value of $V_{ME}$ as a function of frequency of a.c. magnetic field at $H_{AC}$ = 1.3 Oe and $V_{DC}$ = -0.7 V. The arrow indicates the resonance frequency is around 933 Hz. **c**, The maximum value of $V_{ME}$ as a function of magnitude of $H_{AC}$ at $f_r$ = 933 Hz and $V_{DC}$ = -0.7 V. The straight line indicates the linear dependence of $V_{ME}$ with $H_{AC}$.

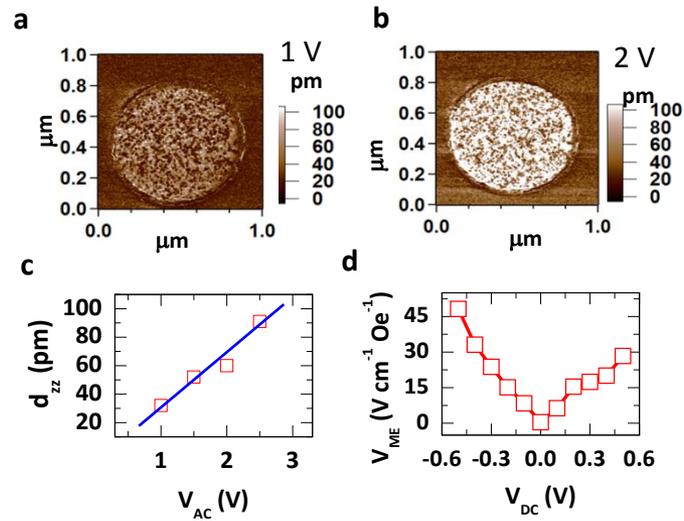

**Figure 3 | Development of strain in a few atomic layers of dielectric MgO under the influence of applied *E*-field.**
**a** and **b**, PFM images of the nanopillar device with structure: bottom electrode/1.2 nm CoFeB/5 nm MgO/10 nm top electrode at a.c. voltage amplitude of 1 V and 2 V. A strong development of strain can be seen at 2 V. **c**, Strain coefficient ($d_{zz}$) values at different a.c. voltage amplitudes. A blue line indicates the linear dependence of $d_{zz}$ with applied voltage. **d**, The peak value of $V_{ME}$ observed as a function of $V_{DC}$.

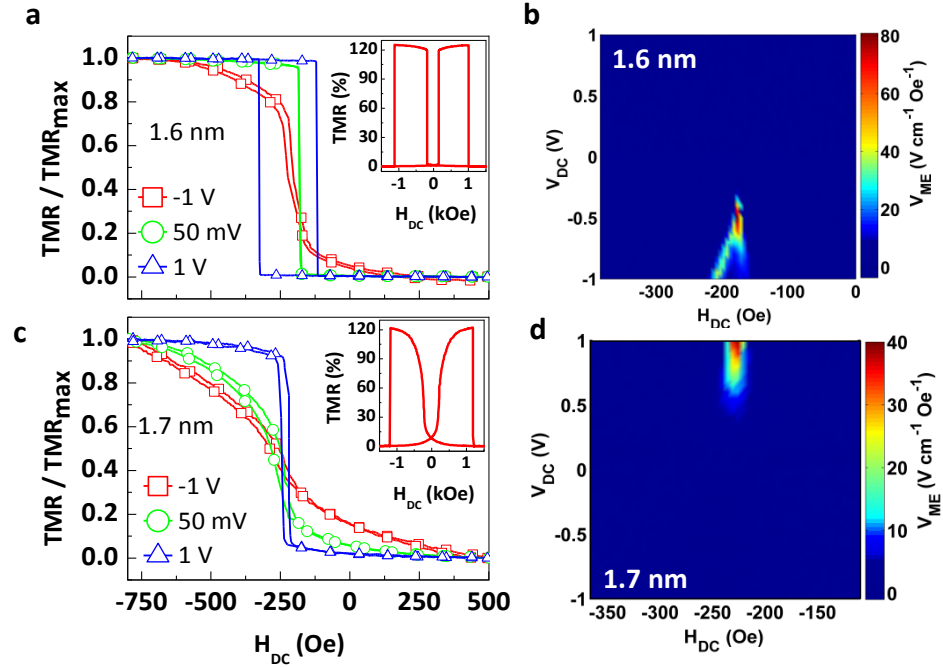

**Figure 4 | The effect of *E*-field on magnetization of thicker CoFeB FL, and its response to ME effect.**
**a**, Normalized minor TMR loops at $V_{DC}$ = 50 mV, -1 V and +1 V for a MTJ device with 1.6 nm CoFeB FL which has PMA at low *E*-field and **b**, shows the $V_{DC}$ and $H_{DC}$ map of *E*-field induced ME voltage for the same device at $f_r$ = 2733 Hz under $H_{AC}$ = 1.3 Oe. **c**, and **d**, show the same for a MTJ device with 1.7 nm CoFeB FL with tilted magnetization at low *E*-field. The insets in Figs. **a**, and **c**, show the full TMR loops at $V_{DC}$ = 50 mV.

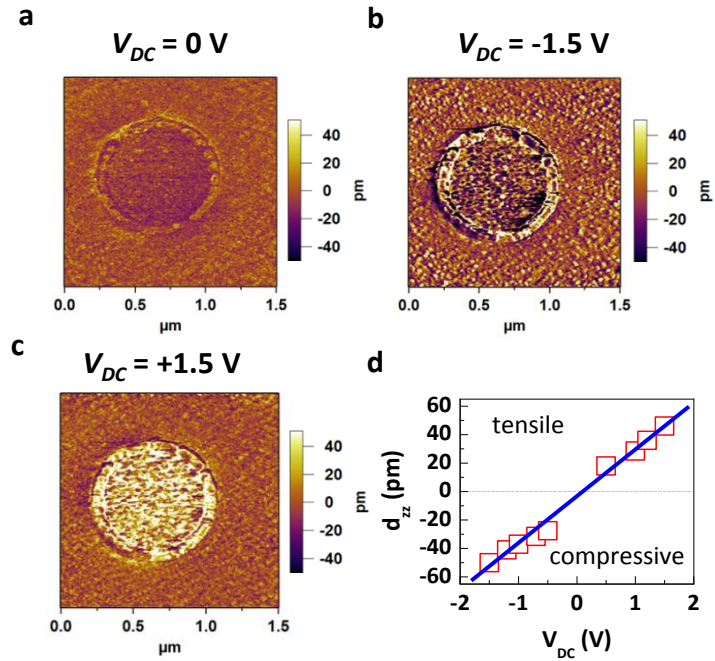

**Figure 5 | Effect of d.c. *E*-field polarity on the piezoresponse of nanoscaled MgO.**
**a**, **b** and **c**, PFM images of nanopillar device with 5 nm MgO under $V_{DC}$ = 0, -1.5 and 1.5 V, respectively. As can be seen, negative $V_{DC}$ leads to compressive strain, while the positive $V_{DC}$ gives rise to tensile strain. **d**, $d_{zz}$ values at different d.c. voltage amplitudes.

# Electric-field-induced strain-mediated magnetoelectric effect in CoFeB-MgO magnetic tunnel junctions

V. B. Naik,[1,*] H. Meng,[1] J. X. Xiao,[2] R. S. Liu,[1] A. Kumar,[2] K. Y. Zeng,[2] P. Luo[1] and S. Yap[1]

[1]Data Storage Institute, A*STAR (Agency for Science Technology and Research), 5 Engineering Drive 1, DSI Building, Singapore 117608, Singapore.

[2]Department of Mechanical Engineering, National University of Singapore, 9 Engineering Drive 1, Singapore 117576, Singapore.

# Supplementary Information

## I. Top CoFeB free layer thickness dependence study

In order to investigate the magnetization switching behavior in the magnetic tunnel junctions (MTJs) devices with different CoFeB free layer (FL) thickness ($t$), we have fabricated the MTJ structures made of Si/SiO$_2$/bottom electrode/Ta 2/CoFeB 1/MgO 1.8/CoFeB $t$ = 1.2-1.7/Ta 5/Ru 10 (numbers are nominal thicknesses in nanometres). The multilayer was processed into circular MTJ devices with a diameter of 85 nm by electron beam lithography (EBL) and ion beam etching processes and annealed in a vacuum oven at 300 °C for 1 hour. The magnetoresistance loops were measured at room temperature with a d.c. voltage ($V_{DC}$) under a perpendicular scanning magnetic field ($H_{DC}$). Here, the positive bias voltage is defined as inducing the tunneling of electrons from the bottom CoFeB electrode to the top CoFeB electrode.

---

[*] Email: Vinayak_BN@dsi.a-star.edu.sg or vinuprl@yahoo.com

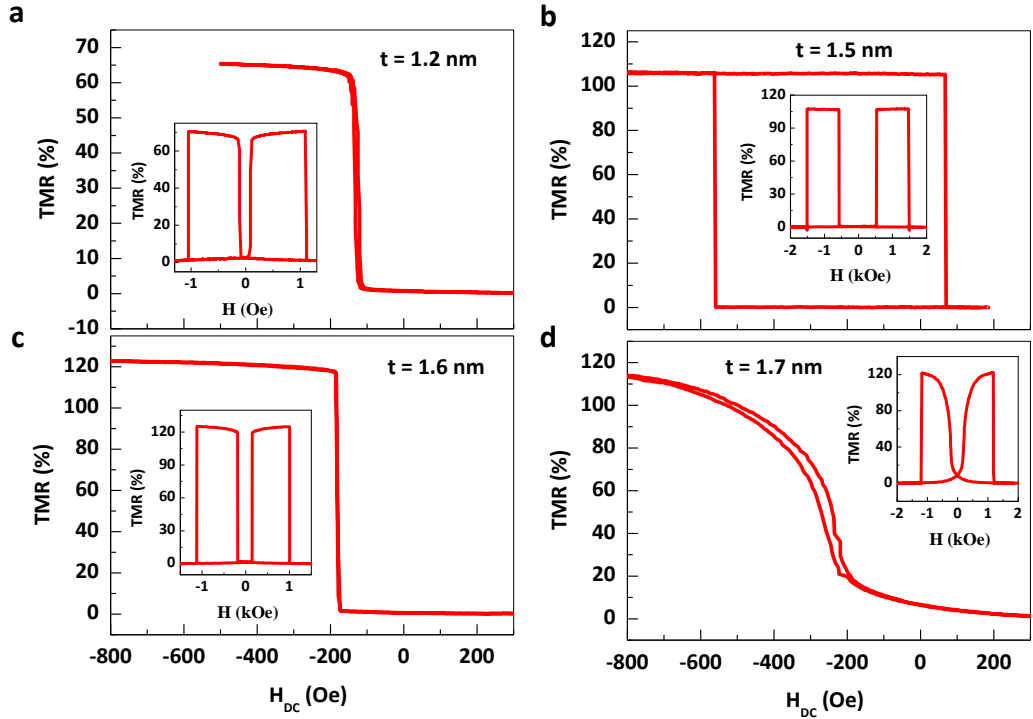

**Figure S1 | Minor TMR loops of the devices with different top CoFeB thicknesses (*t*) at $V_{DC}$ = 50 mV. The inset shows the full TMR loop.**

Figure S1 show the minor TMR loops of the devices with different top CoFeB thicknesses: *t* = 1.2, 1.5, 1.6 and 1.7 (numbers are nominal thicknesses in nanometres) at $V_{DC}$ = 50 mV. The inset in all the figures shows the corresponding full TMR loop. Clearly, the bottom CoFeB is the reference layer (RL) with larger switching field and the top CoFeB layer is the free layer (FL) with smaller switching field. The FL magnetization switch is smooth for *t* = 1.2 nm (Fig. S1a), which is due to weak perpendicular magnetic anisotropy (PMA). However, a sharp square minor TMR loop is observed for *t* = 1.5 nm with a sufficiently large coercive field ($H_C$) of 315 Oe (Fig. S1b). A magnetic easy axis changes from perpendicular to in-plane when the thickness of FL is increased from 1.6 nm (Fig. S1c) to 1.7 nm (Fig. S1d). The evolution of $H_C$ for the FL is due the change from PMA to in-plane magnetic anisotropy when the thickness of FL increases.

## II. Electric-field-controlled PMA in MTJ devices with different free layer thicknesses

In order to investigate the effect of electric-field ($E$-field) on the PMA properties of FL with different thicknesses, we have measured minor TMR loops of the devices with $t$ = 1.2-1.7 nm at different applied d.c. bias voltages ($V_{DC}$ = -1 V to 1 V).

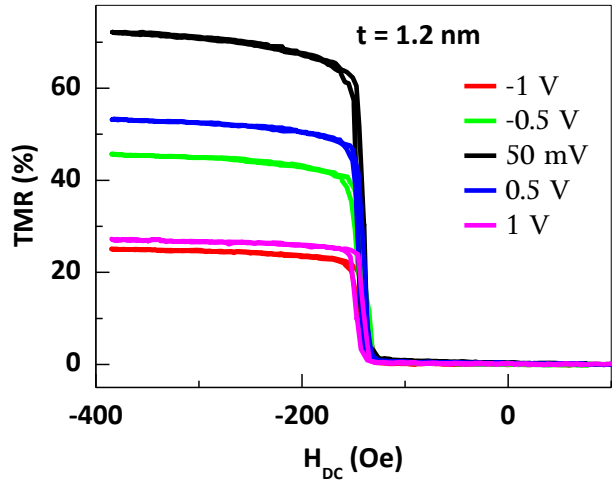

**Figure S2 | Minor TMR loops of junction with 1.2 nm top CoFeB FL at different applied d.c. bias voltages (-1 V to 1 V).**

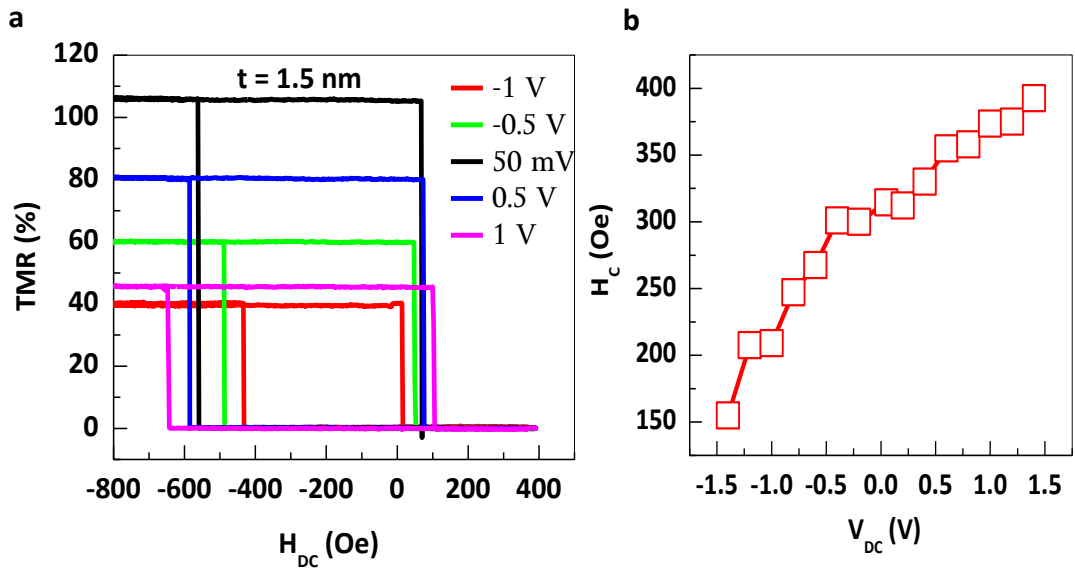

**Figure S3 | a, Minor TMR loops of junction with 1.5 nm top CoFeB FL at different applied d.c. bias voltages (-1 V to 1 V). b, The coercive field ($H_C$) as a function of applied d.c. bias voltage.**

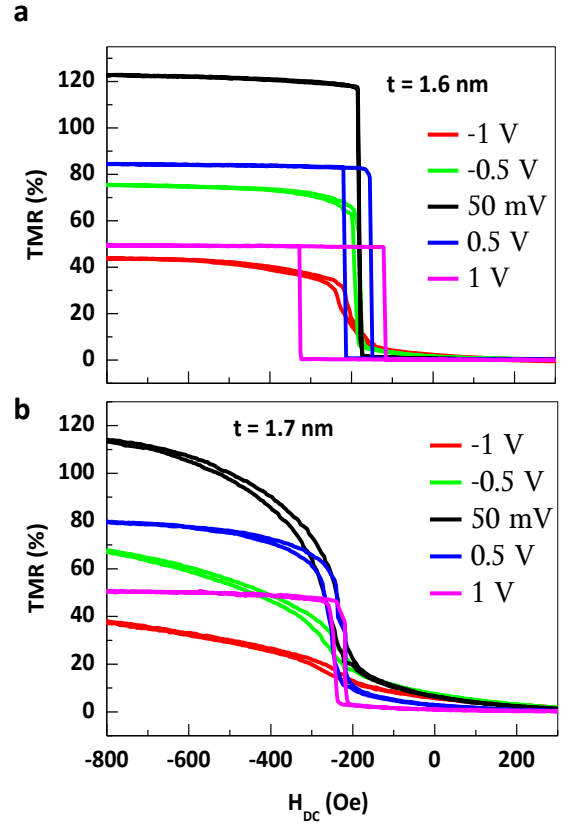

**Figure S4 | Minor TMR loops of junctions with (a) 1.6 nm and (b) 1.7 nm top CoFeB FL at different applied d.c. bias voltages (-1 V to 1 V).**

Fig. S2 shows the minor TMR loops for $t = 1.2$ nm at different $V_{DC}$, which reveals that there is not obvious change of PMA even under a large bias voltage for such a thin FL. A strong $E$-field-controlled PMA change is observed for $t = 1.5$ nm (Fig. S3a), and the PMA becomes stronger under positive bias voltages ($V_{DC} > 0$), while it reduces under negative bias voltages ($V_{DC} < 0$ V). The increased $H_C$ at positive $V_{DC}$ and decreased $H_C$ at negative $V_{DC}$ (Fig. S3b) are consistent with the previous reports[1,2,3,4]. The $H_C$ is almost zero under $V_{DC} = 50$ mV for $t = 1.6$ nm, at which the magnetic easy axis is on the verge of changing from perpendicular to in-plane (Fig. S4a). On the application of positive $V_{DC}$, $H_C$ attains a non-zero value and increased up to 105 Oe at $V_{DC} = 1$ V. On the other hand, the application of negative $V_{DC}$ changes the magnetic easy axis from perpendicular to in-plane. In-plane magnetic easy axis is observed for $t = 1.7$ nm (minor TMR loop at $V_{DC} = 50$ mV in Fig. 4b). As expected, the positive $V_{DC}$ pulls the magnetic easy axis to perpendicular from in-plane, while the negative $V_{DC}$ makes it more in-plane.